\documentclass{article}
\usepackage{color}
\usepackage{spconf,amsmath,graphicx, multirow}
\usepackage{graphicx}

\newcommand{\blue}[1]{\textcolor{black}{#1}}

\title{Beware of diffusion models for synthesizing medical images - \\A comparison with GANs in terms of memorizing brain MRI and chest x-ray images}

\name{Muhammad Usman Akbar$^{\: a,c}$, Wuhao Wang$^{\: b}$, Anders Eklund$^{\: a,b,c}$}

\address{$^a$ Division of Medical Informatics, Department of Biomedical Engineering\\  $^b$ Division of Statistics \& Machine Learning, Department of Computer and Information Science\\ $^c$ Center for Medical Image Science and Visualization (CMIV) \\Link\"{o}ping University, Sweden \\ }
\usepackage[backend=bibtex]{biblatex}
\begin{document}
\maketitle

\begin{abstract}
   Diffusion models were initially developed for text-to-image generation and are now being utilized to generate high quality synthetic images. Preceded by GANs, diffusion models have shown impressive results using various evaluation metrics. However, commonly used metrics such as FID and IS are not suitable for determining whether diffusion models are simply reproducing the training images. Here we train StyleGAN and a diffusion model, using  BRATS20, BRATS21 and a chest x-ray pneumonia dataset, to synthesize brain MRI and chest x-ray images, and measure the correlation between the synthetic images and all training images. Our results show that diffusion models are more likely to memorize the training images, compared to StyleGAN, especially for small datasets and when using 2D slices from 3D volumes. Researchers should be careful when using diffusion models \blue{(and to some extent GANs)} for medical imaging, if the final goal is to share the synthetic images.   
\end{abstract}

\section{Introduction}
            
Synthetic \blue{medical} data have become a very popular research topic recently~\cite{rankin2020reliability,el2021evaluating,yi2019generative,rajotte2022synthetic}, \blue{especially in the medical imaging domain where obtaining large datasets for deep learning is difficult due to different regulations such as GDPR (and anonymizing images is more difficult compared to anonymizing tabular data). Synthetic medical images have the potential to facilitate data sharing~\cite{beaulieu2019privacy,akbar2023brain}, but discussions with legal experts require quantification of memorization and data leakage. Here we therefore investigate the degree of memorization for different generative models. }

Generative adversarial networks (GANs)~\cite{goodfellow2014generative} have for almost a decade been used to generate images with higher and higher resolution. The machine learning community has recently paid a lot of attention to the powerful and adaptable framework known as diffusion models~\cite{ho2020denoising}, which is another type of generative model inspired by anisotropic diffusion for image denoising. In the context of generative models, diffusion refers to the gradual transformation of random noise into meaningful data, such as images or text, through a series of carefully designed steps. The sampling of realistic samples from complex, high-dimensional distributions is the primary objective of both GANs and diffusion models, and diffusion models have been shown to outperform GANs~\cite{dhariwal2021diffusion}. In this work we focus on the topic of memorization, i.e. that generative models can simply generate samples that are copies of the training data.




\subsection{GANs}
    The development of generative image models has a rich past. It was the emergence of GANs~\cite{goodfellow2014generative} that first made it possible to produce high-quality images on a large scale~\cite{karras2019style,brock2018large}. GANs were groundbreaking in deep learning, and allows the creation of realistic and high-quality data, such as images, text, and sound. GANs have two parts: a generator and a discriminator. The generator makes synthetic data, and the discriminator determines if the data is real or synthetic. The generator and discriminator work in competition to each other to produce more and more realistic data. StyleGAN~\cite{karras2019style} is a special kind of GAN that is famous for making detailed and clear images, while giving great control over the final result.
    

\subsection{Diffusion models}
    In the past couple of years, diffusion models~\cite{ho2020denoising} have primarily overtaken GANs, achieving top-notch performance in academic benchmarks~\cite{dhariwal2021diffusion} and serving as the foundation for widely-recognized image generators such as Stable Diffusion~\cite{rombach2022high}, DALL-E 2~\cite{ramesh2021zero,ramesh2022hierarchical}, Imagen~\cite{saharia2022photorealistic} and others. One can say that diffusion probabilistic models are conceptually rather simple as they can be classified as image denoisers \cite{ho2020denoising}. Despite their primary objective being to remove noise, diffusion models are capable of generating high-quality images from random noise. This denoising process occurs incrementally rather than in a single step. Diffusion models operate iteratively, progressively changing the noise to produce a refined image. This is achieved by training the model with varying degrees of noise, enabling it to adapt and produce high-quality results.

\begin{table*}[htb]
\begin{center}
\begin{tabular}{|c|c|c|c|c|}
\hline
\textbf{Paper}              & Number of training subjects and/or images & Dimension & Modality & Mentions memorization\\ \hline

Ali et al. ~\cite{ali2023spot} & 3165 images & 2D & chest x-ray & No \\

Moghadam et al.~\cite{moghadam2023morphology} & 344 (33,777 patches) & 2D & histology & No \\

Pan et al.~\cite{pan20232d} & 5500 images &  2D  & chest x-ray & No\\ 

Pan et al.~\cite{pan20232d} & 101 (1902 images) &  2D  & heart MRI & No \\ 

Pan et al.~\cite{pan20232d} & 94 (4157 images) &  2D  & pelvic CT & No \\ 

Pan et al.~\cite{pan20232d} & 30 (2178 images) &  2D  & abdomen CT & No \\ 

Peng et al.~\cite{peng20232d} & 3000 images &  2D  & brain MRI & No \\

Txurio et al.~\cite{txurio2023diffusion} & 447 (57,216 images) &  2D  & pelvic CT & No \\

Packhäuser et al.~\cite{packhauser2022generation} & 30,805 (112,120 images) & 2D  & chest x-ray & Yes \\ 

Pinaya et al.~\cite{pinaya2022brain} & 31,740 & 3D &  brain MRI & No \\  

Khader et al.~\cite{khader2023denoising} & 998 & 3D & brain MRI & No \\

Khader et al.~\cite{khader2023denoising} & 1010 & 3D & thoracic CT & No \\

Khader et al.~\cite{khader2023denoising} & 1844 & 3D & breast MRI & No \\

Khader et al.~\cite{khader2023denoising} & 1250 & 3D & knee MRI & No \\

Dar et al.~\cite{dar2023investigating} & 65 (300 sub-volumes) & 3D & coronary CTA & Yes \\

Dar et al.~\cite{dar2023investigating} & 904 & 3D & knee MRI & Yes \\

\hline

\end{tabular}%
\caption{A comparison of previous papers which used diffusion models to synthesize medical images from noise. For papers which synthesize several types of images or volumes, we list one row per modality. There is a wide range in the number of subjects (or images) used for training the diffusion models. Based on our results, diffusion models trained with a small number of subjects or images are likely to memorize the training data.}
\label{tab:previousworktable}
\end{center}
\end{table*}
\subsection{Evaluation metrics}

How to evaluate the quality and diversity of synthetic images is an open research question and many different metrics have been proposed~\cite{borji2019pros,borji2022pros}. \blue{Borji~\cite{borji2019pros} review and critically discuss more than 24 quantitative and 5 qualitative metrics, including the most common metrics Frechet inception distance (FID)~\cite{heusel2017gans} and inception score (IS)~\cite{salimans2016improved}. Another kind of evaluation is to train a classifier or a segmentation network using the synthetic images~\cite{eilertsen, akbar2023brain}, and then test the networks on real images. This approach will better reflect how useful the synthetic images are, but it can be very time consuming if many generative models have been trained.} Many metrics, such as FID and IS, do not consider memorization or overfitting. A possible explanation for this is that memorization has in general not been a problem for GANs.


\subsection{Diffusion models in medical imaging}

In medical imaging, diffusion models are still relatively rare; GANs are still the most common architecture for noise-to-image and image-to-image generation~\cite{yi2019generative}. Some notable exceptions are covered here, see Table~\ref{tab:previousworktable} for an overview. Pan et al.~\cite{pan20232d} implemented a diffusion model with transformer based denoising, instead of the common U-Net based approach, and used it to generate chest x-rays, heart MRI, pelvic CT, and abdomen CT. For a COVID-19 classification task, the baseline network obtained an accuracy of 0.88 using a pure real dataset, 0.89 using a pure synthetic dataset, and 0.93 using a dataset mixed of real and synthetic data. Pinaya et al.~\cite{pinaya2022brain} implemented a conditional 3D diffusion model to synthesize brain volumes, and was able to control for example age and ventricular volume. However, the synthetic volumes were not used to train any model. Moghadam et al.~\cite{moghadam2023morphology} used diffusion models to synthesize high quality histopathology images of brain cancer, and demonstrated better IS and FID metrics compared to a GAN. Ali et al.~\cite{ali2023spot} used diffusion models to generate lungs X-Ray and CT images, and then let radiologists evaluate the images as real or synthetic. Packhäuser et al.~\cite{packhauser2022generation} used diffusion models to generate chest radiographs for training thoracic abnormality classification systems, and demonstrated that diffusion models outperform GANs. Khader et al~\cite{khader2023denoising} used 3D diffusion models to synthesize brain MRI, thoracic CT, breast MRI as well as knee MRI volumes. The volumes were evaluated by radiologists, and by pre-training segmentation models. Kazerouni et al.~\cite{kazerouni2022diffusion} provide a survey of diffusion models in medical imaging, but the number of papers on noise-to-image diffusion models is rather low (most are mentioned in this paper). 

\subsection{Memorization}

\blue{Both GANs and diffusion models represent advanced generative approaches, but they differ fundamentally in their training methods and mechanisms of generating new data, which impacts how they "remember" training data and why diffusion models, in particular, require careful handling.
GANs are trained through a competitive process between two neural networks: a generator that creates images and a discriminator that evaluates them, through an adversarial loss function. The generator only improves its output based on feedback from the discriminator, i.e. without seeing the training images, aiming to produce images indistinguishable from real examples. The fact that the generator never sees the training images will limit memorization. Diffusion models, on the other hand, operate through a process of adding and then iteratively removing noise from a data sample. They start by gradually adding noise to an image until the original is completely obscured, and then they learn to reverse this process to generate new images from pure noise. Because this process involves learning a very detailed noise reduction path, for example by minimizing the mean squared error between original and denoised images, it inherently incorporates detailed characteristics from the training data at each step. Thus unlike GANs, diffusion models do not only learn the general structure of the data, but can potentially learn specific instances making it more prune to learning the training data.}

Surprisingly, most of the papers in Table~\ref{tab:previousworktable} do not mention memorization or overfitting. Pan et al.~\cite{pan20232d} calculated a diversity score (DS) to obtain a measure of the diversity of the synthetic images. For each synthetic image, they calculated the structural similarity index measure (SSIM) between it and every other synthetic image to find the most similar pair. This procedure was repeated for all pairs of real images, and the distributions of nearest synthetic and real SSIM were compared using Kullback-Leibler divergence. Unfortunately the authors did not calculate SSIM between each synthetic image and all training images. To investigate memorization is even more important for larger models; their diffusion model had 1.6 billion trainable parameters (which is substantially higher compared to the GANs used for comparison). Pinaya et al.~\cite{pinaya2022brain} also used SSIM to investigate the diversity of the generated brain volumes, but did not calculate SSIM between real and synthetic volumes. Khader et al.~\cite{khader2023denoising} also used SSIM to investigate the diversity of the synthetic volumes, but did not use SSIM to detect memorization. Packhäuser et al.~\cite{packhauser2022generation} guaranteed patient privacy by using additional networks to remove synthetic images that are too similar to the training images. However, it is not mentioned how many synthetic images that were seen as too similar. The authors used 112,120 chest radiographs from 30,805 patients for training, and for such a large dataset it is less likely that a diffusion model can memorize a large proportion of the images. Kazerouni et al.~\cite{kazerouni2022diffusion} only mention privacy in a short paragraph in their survey. Dar et al.~\cite{dar2023investigating} investigated memorization for 3D diffusion models, but did not compare with a 3D GAN and did not investigate the effect of dataset size.

\subsection{Related work on memorization}

Here we compare a 2D GAN and a 2D diffusion model in terms of memorization, for synthetic brain MRI and chest x-ray images, and demonstrate that diffusion models are much more prone to memorizing the training images. This can be very problematic in medical imaging where privacy is of uttermost importance. In our previous work on training segmentation networks with synthetic images~\cite{larsson2022does,akbar2023brain} the obtained Dice scores were used for evaluation. When using synthetic images from diffusion models, it was observed that the Dice scores were as high as when training with real images, which made us suspicious. 

\blue{This work can be seen as a continuation of several previous studies. Carlini et al.~\cite{carlini2023extracting} and Somepalli et al.~\cite{somepalli2023diffusion} investigated memorization in diffusion models, but their work did not focus on medical images. Pinaya et al.~\cite{pinaya2022brain} and Khader et al.~\cite{khader2023denoising} used 3D diffusion models to synthesize brain volumes but did not address memorization. Dar et al.~\cite{dar2023investigating} examined memorization in 3D diffusion models but did not compare these findings with 3D GANs. Packhäuser et al.~\cite{packhauser2022generation} removed overly similar synthetic images but did not report the significance of the memorization problem.}

\section{Methods}
\blue{This section describes the two distinct datasets used (brain MRIs and chest x-rays), as well as the two generative models (StyleGAN and a diffusion model). Furthermore it describes different metrics and approaches for quantifying memorization. Finally, it explains how we tried different hyperparameters for the generative models, to see how they affect the degree of memorization.}


\subsection{Data}
\subsubsection{ Brain tumor MRI }

\blue{Deep learning can be used for many tasks in brain MRI, such as brain disease diagnosis, brain age prediction, brain anomaly detection and segmentation~\cite{chen2023understanding}, but all tasks require large datasets for training.} The MR images used for this project were downloaded from the multimodal brain tmour segmentation Challenge (BRATS)~\cite{bakas3,bakas4,bakas1,bakas2,menze}. The dataset contains MR volumes of size 240 × 240 × 155, and for each patient four types of MR images are available: T1-weighted, post gadolinium contrast T1-weighted, T2-weighted, and T2 fluid attenuated inversion recovery (FLAIR). The annotations cover three parts of the brain tumor: peritumoural edema (ED), necrotic and non-enhancing tumour core (NCR/NET), and GD-enhancing tumour (ET). Two datasets were used for our experiments, BRATS20 (369 subjects) and BRATS21 (1251 subjects). The large difference in the number of subjects is excellent for testing how the memorization depends on the number of training images (which was not investigated by Carlini et al.~\cite{carlini2023extracting}). For each dataset, 56 subjects were reserved for testing.

All 3D volumes were split into 2D slices, as a 2D GAN and a 2D diffusion model were used (3D GANs and 3D diffusion models are not yet very common). Only slices with at least 15\% pixels with an intensity of more than 50 were included in the training. This resulted in a total of 23,478 5-channel images for BRATS20, and 91,271 5-channel images for BRATS21. Each slice was zero padded from 240 x 240 to 256 x 256 pixels, as the used GAN only works for resolutions that are a power of 2, and the intensity was rescaled to 0 - 255. The intensities for the tumor annotations were changed from [1,2,4] to [51,102,204], such that the intensity range is more similar for the 5 channels. Figure~\ref{fig:Samples_Brats_synthetic} shows example images from the BRATS20 and BRATS21 datasets, displaying all four modalities and their corresponding segmentation masks.

\subsubsection{Chest x-ray}

\blue{Deep learning can for chest x-rays be used for image-level prediction (classification and regression), segmentation, and localization~\cite{ccalli2021deep}, but requires large datasets for training.}
For comparison with 2D data, we therefore used the "Chest X-Ray Images (Pneumonia)" dataset from Kaggle~\cite{kermany2018identifying}. This dataset consists of anterior-posterior chest X-ray (CXR) images of pediatric patients aged between one to five years. For suitable alignment during GAN training, the images, which originally varied in dimensions with an average size of 1320 x 968 pixels, were resized to a uniform size of 256 x 256 pixels. The dataset comprises 5,863 X-ray images in JPEG format, split into two categories: pneumonia and normal. Of these, 5,216 images were allocated for training, while the remaining 624 were set aside for testing. Due to a discrepancy in the number of images between BRATS20 and BRATS21, we also employed a subset of 1,300 CXR images to maintain a similar ratio. This approach facilitated a balanced comparison between CXR-5216 and CXR-1300, mirroring the comparison between BRATS20 and BRATS21.


\subsection{Image generation}

 The StyleGAN architecture~\cite{karras2019style} was used for all GAN experiments, starting with the default settings. For BRATS data the code was modified to generate 5-channel (T1w, T1wGD, T2w, FLAIR, segmentation annotations) images instead of 3-channel color images. The GAN will thereby learn to jointly generate 4-channel MR images and the corresponding tumor annotations at the same time. For the pneumonia data the code was modified to one channel instead of three channel color images. For each dataset, the training took 3 days on 8 Nvidia A100 GPUs.

The diffusion model proposed by Ho et al.~\cite{ho2020denoising} was used for all diffusion experiments, starting with the default settings. For BRATS data the code was modified to generate 5-channel images instead of 3-channel color images. Training the diffusion model took 22 days on single Tesla V100 GPU. For pneumonia data the code was modified to cater a one-channel grey scale image. The training took 6 days on a single Tesla V100 GPU
 

\begin{figure*}[h!]
  \centering
  \includegraphics[width=0.96\textwidth]{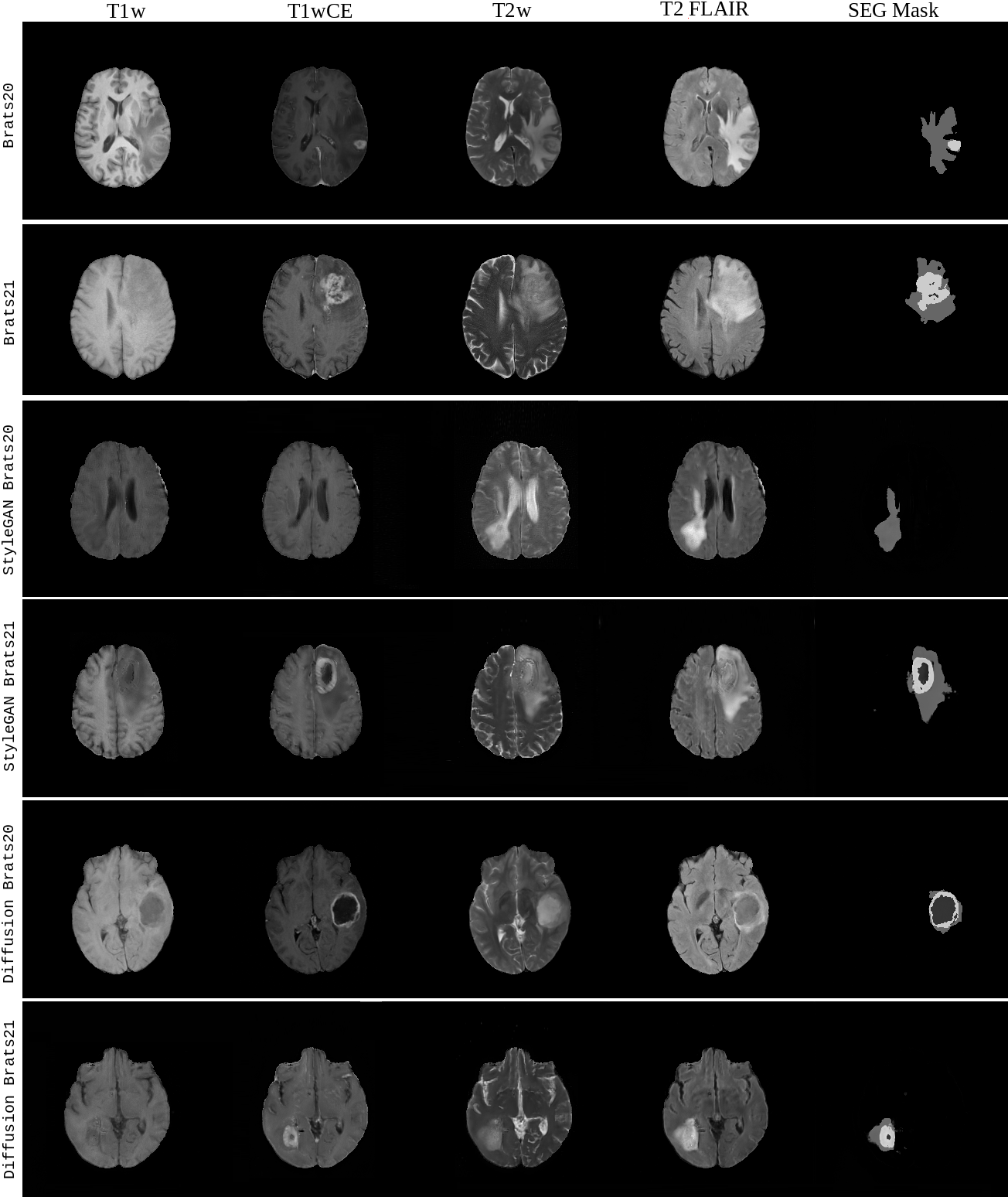} 
  \caption{Row 1: Sample 5-channel image from BRATS20 data. Row 2: Sample 5-channel image from BRATS21 data. Row 3: synthetic 5-channel image from StyleGAN trained with BRATS20. Row 4: synthetic 5-channel image from StyleGAN trained with BRATS21. Row 5: synthetic 5-channel image from diffusion model trained with BRATS20. Row 6: synthetic 5-channel image from diffusion model trained with BRATS21.}
  \label{fig:Samples_Brats_synthetic}
\end{figure*}

\subsection{Similarity metrics for memorization}

\blue{Different similarity metrics can be used to detect memorization of the training images.} Image correlation is an important concept in the field of computer vision and image processing. It is a quantitative measure that gauges the similarity or relationship between two images, and can be used to study memorization. \blue{It is also an effective metric for detecting memorization in generative models because it directly compares individual pixel values between generated and training images, with high correlation indicating potential memorization. Its sensitivity to fine details allows it to capture subtle similarities, providing a clear and quantitative measure of resemblance. Additionally, pixel-wise correlation is straightforward to compute and interpret, enabling quick assessments of potential memorization without the need for complex methods.}

\blue{Consider a scenario with 1000 generated images. Each generated image is compared with every original image in the training set, as it is sufficient if a single training image is memorized. For the BRATS20 dataset, which contains 23,478 training images, and the BRATS21 dataset, which contains 91,271 training images, this results in a total of 23,478,000 (BRATS20) + 91,271,000 (BRATS21) = 114,749,000 comparisons to investigate memorization. For the CXR data, where the training set comprises 5216 and 1300 images, the total number of comparisons is 5,216,000 + 1,300,000 = 6,516,000. This extensive comparison process ensures that even the most subtle similarities between the generated and training images are detected, providing a robust measure for identifying memorization in generative models. However, it also makes the process computationally expensive, making it challenging to add more advanced metrics due to the sheer number of comparisons required.}

\blue{Other metrics, such as mutual information (MI) and structural similarity index (SSIM), can also be used to detect memorization. Mutual information measures the amount of information shared between images, and SSIM assesses the perceived quality of images by comparing structural information. However, these metrics are more computationally demanding compared to correlation (a small comparison in Matlab revealed that SSIM is more than 10 times as demanding to calculate, and MI is 50 times as demanding as it first requires estimation of a joint histogram). Although these metrics might provide additional insights into memorization, their higher computational cost makes them less feasible for large-scale comparisons.}

\blue{Another approach for quantifying memorization could be to use the feature vector in an autoencoder, which can work effectively even for rotated synthetic images for quantifying the memorization. An autoencoder maps images to a lower-dimensional feature space, typically between 100 to 1000 dimensions. While this approach is computationally efficient, the reduced dimensionality might not capture the fine details necessary for detecting memorization as effectively as pixel-wise correlation. Thus, the feature vector approach might not be as sensitive in identifying subtle memorization in generative models.}

\subsection{How do hyperparameters influence memorization?}

\blue{In deep learning the number of trainable parameters a model has is strongly related to how likely the model is to overfit (memorize). For chest x-ray images, StyleGAN has a total of 
57,335,748 trainable parameters (generator: 28,471,875, discriminator: 28,863,873), while the diffusion model has 
31,050,885 trainable parameters (we used the default settings for both models).} \blue{For our brain MR images, which have 5 channels, StyleGAN has 
59,962,308 trainable parameters (generator: 31,098,435, discriminator: 28,863,873) and the diffusion model has 41,123,331.}

\blue{In an attempt to study how the size of the generative models affect memorization, a smaller and a larger version of StyleGAN and the diffusion model were therefore trained using the small datasets. For StyleGAN the number of maximum feature maps and the number of capacity multipliers were changed to lower or increase the number of trainable parameters. For the diffusion model the number of channels (filters) was changed. All other hyperparameters remained the same.} \blue{For x-ray images the number of trainable parameters for diffusion models are 7,780,933 and 69,809,861 for smaller and larger versions respectively. For StyleGAN the number of trainable parameters are 47,542,337 (generator: 23,455,739 , discriminator: 24,086,598) and 76,583,792 (generator: 40,260,785 , discriminator: 36,323,007) respectively for smaller and larger versions. Using brain MR data the number of trainable parameters are for the diffusion model 10,304,259 and 92,457,219, for smaller and larger models respectively. For StyleGAN the number of trainable parameters are 15,581,971 (generator: 8,343,757, discriminator: 7,238,214) and 220,400,166 (generator: 105,533,217, discriminator: 114,868,977) respectively for smaller and larger models}

\blue{The resulting synthetic images had random artefacts (not present for the default settings), and the training time was for this reason increased for training the smaller and larger models.}

\section{Results and Analysis} 

\begin{figure*}[h!]
  \centering
  \includegraphics[width=0.99\textwidth]{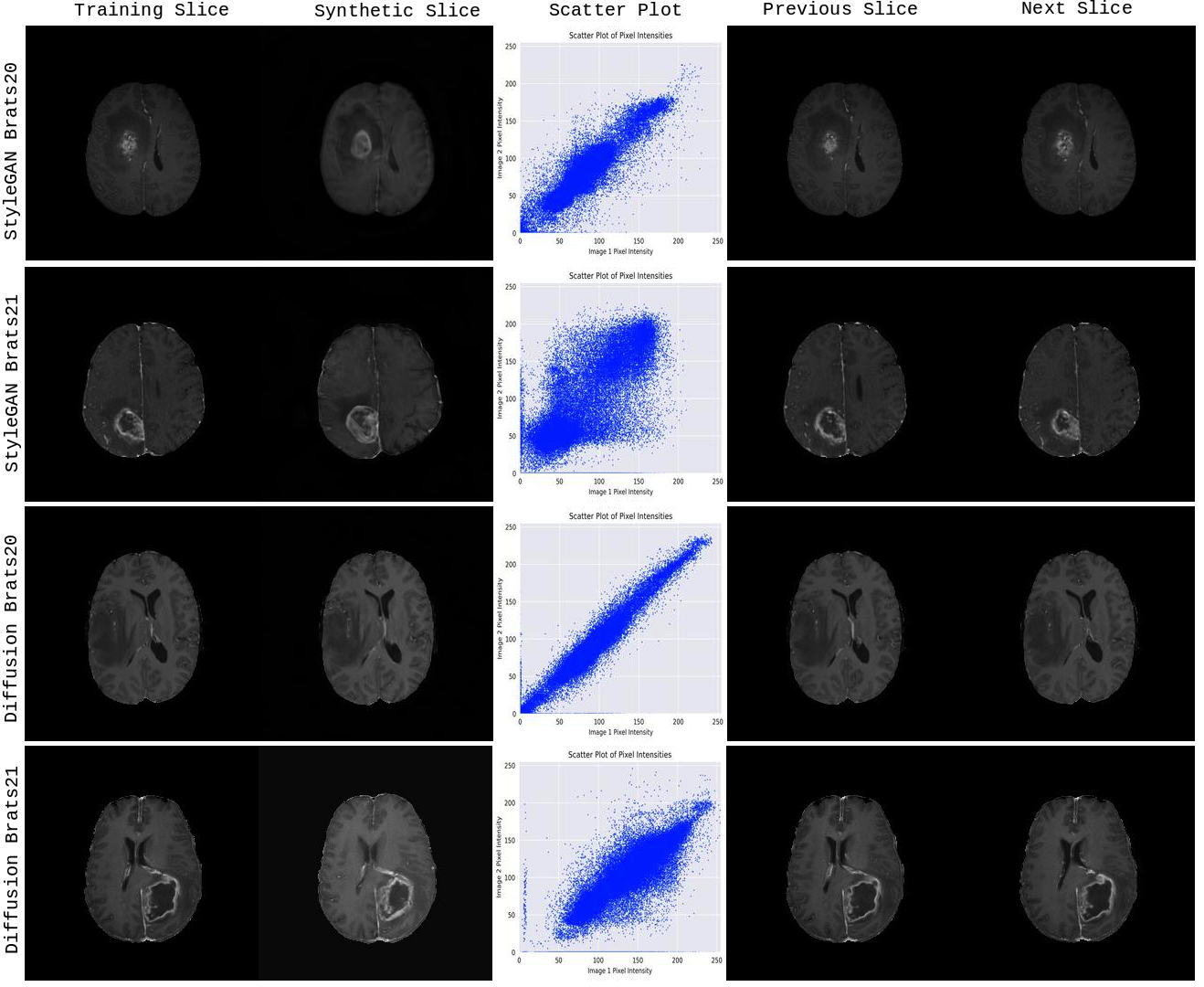}

  \caption{A comparison of randomly selected synthetic images and the training image with the highest correlation. The first column displays the T1wGd training slice with the highest correlation, while the second column presents the corresponding synthesized image. The third column shows a scatter plot comparing the two images. The fourth column features the previous adjacent slice of the training T1wGd image, and the fifth and final column displays the next adjacent T1wGd slice of the training image. 
  Rowwise the first row showcases samples from StyleGAN trained on BRATS20 \blue{with a correlation of 0.96247}, and the second row displays samples trained on BRATS21 \blue{with a correlation of 0.92620}. The third row presents samples from a diffusion model trained on BRATS20 \blue{with a correlation of 0.99002}, while the last row exhibits samples from a diffusion model trained on BRATS21 \blue{with a correlation of 0.97655}. The synthetic images from the diffusion model are more or less copies of a specific training image.}
  \label{fig:braincomparison}
\end{figure*}

\begin{figure*}[h!]
  \centering
  \includegraphics[width=0.8\textwidth]{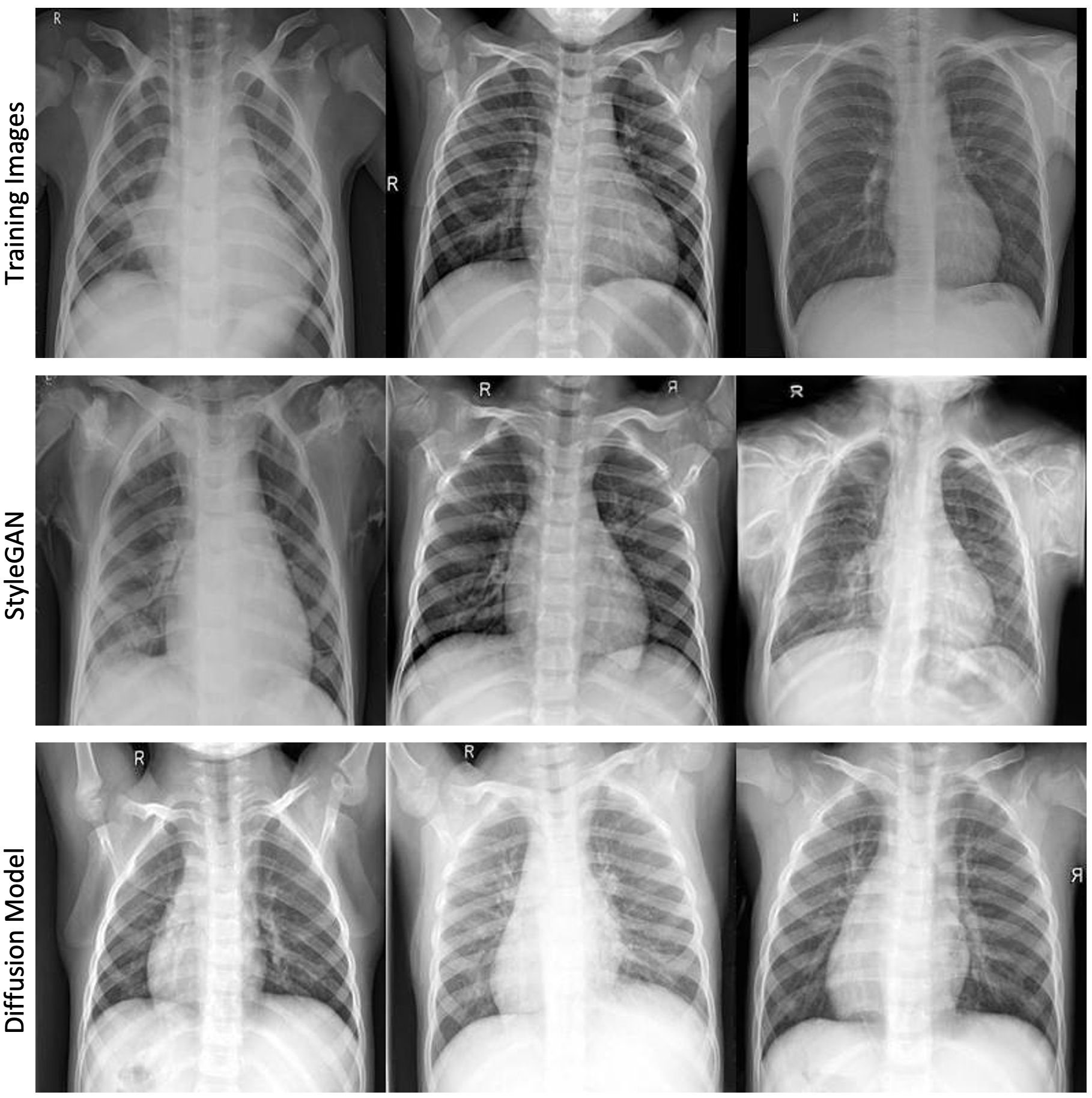} 
  \caption{A comparison of real and synthetic images. Row 1 presents samples from the training data. Row 2 showcases synthetic images generated using StyleGAN. Row 3 depicts synthetic images produced by the diffusion model.}
  \label{fig:Sample_CXR}
\end{figure*}

\begin{figure*}[h!]
  \centering
  \includegraphics[width=0.85\textwidth]{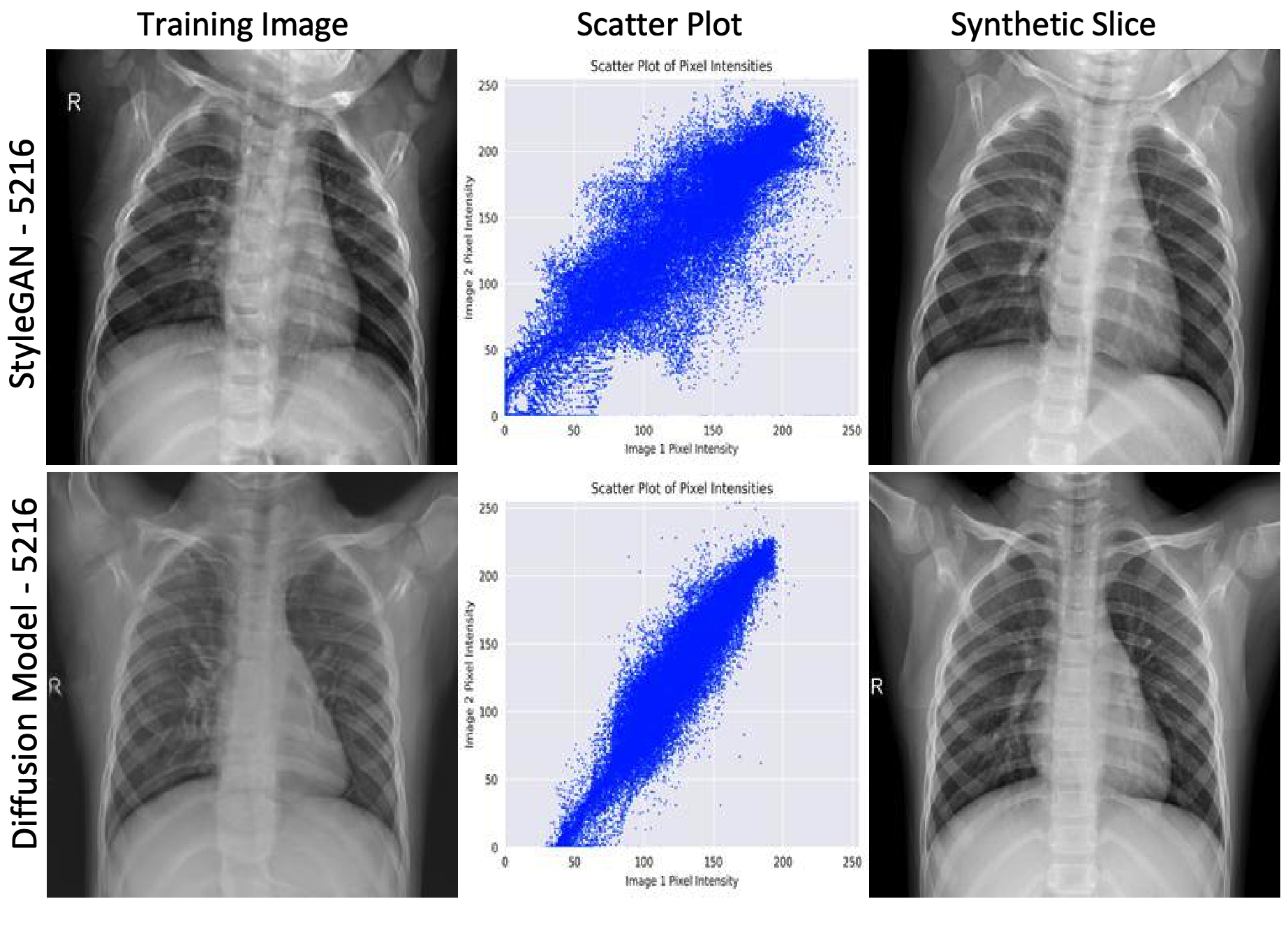} 
  \includegraphics[width=0.85\textwidth]{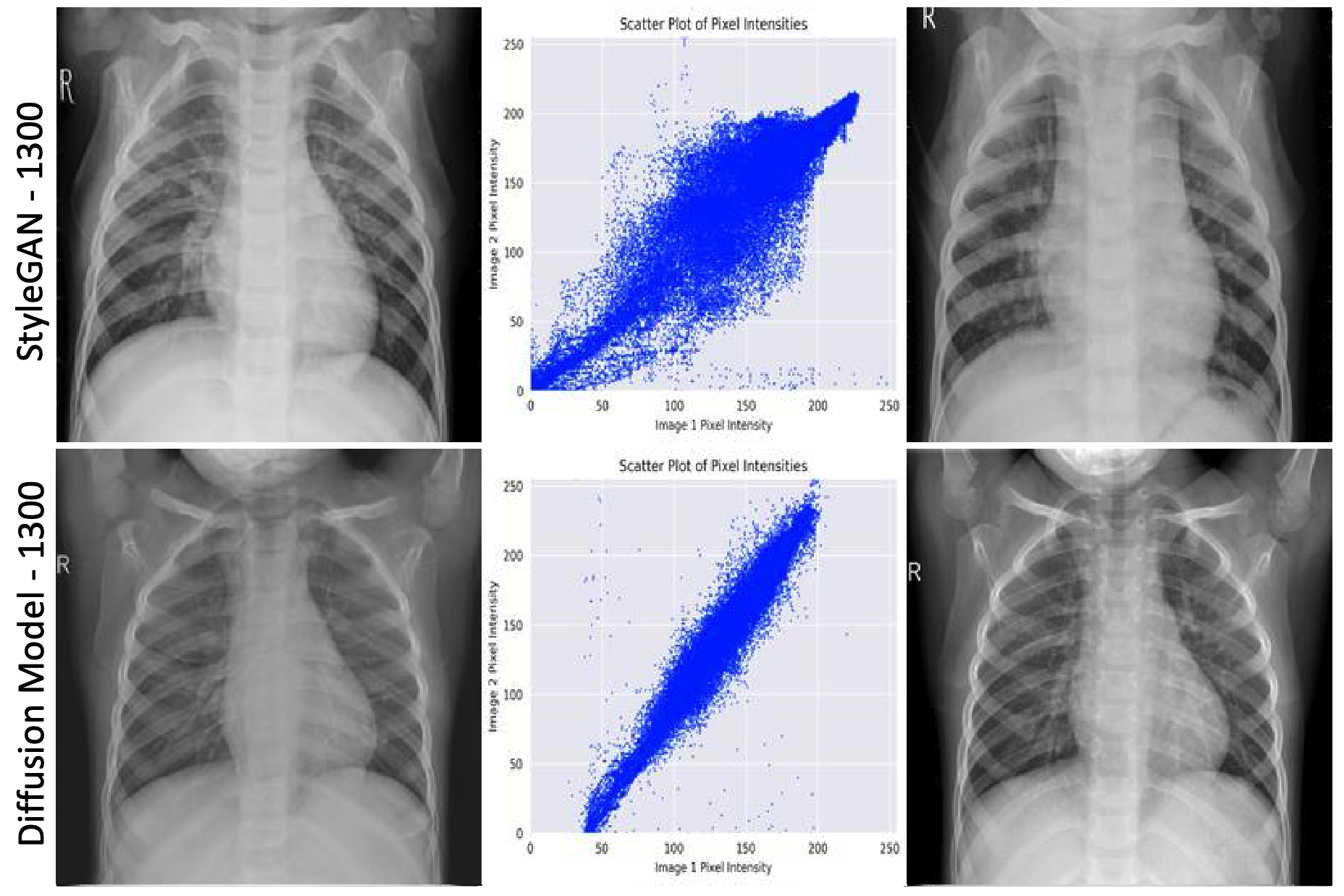} 
  \caption{Comparison between synthetic images and their closest training image.
  Row 1 displays a StyleGAN-generated image and the training image with the highest \blue{correlation of 0.87679} from the CXR-5216 dataset.
  Row 2 showcases an image produced by a diffusion model, also trained on the CXR-5216 dataset, \blue{with a correlation of 0.90564}, alongside the closest training image.
  Rows 3 and 4 show the same comparison using the smaller CXR-1300 dataset, \blue{with a correlation of 0.89733 for the StyleGAN-generated image and 0.92314 for the diffusion model image}.
  Accompanying these images are scatter plots for each of the four rows, providing a visual representation of the correlations.
}
  \label{fig:Coorelation_CXR}
\end{figure*}

In Figures~\ref{fig:Samples_Brats_synthetic} and~\ref{fig:Sample_CXR}, we can see sample training images that were used to train the diffusion model and StyleGAN, along with randomly selected synthetic images from each model. \blue{Figure~\ref{fig:braincomparison} compares synthetic images generated by StyleGAN and a diffusion model trained on BRATS20 and BRATS21 datasets while Figure~\ref{fig:Coorelation_CXR} compares images generated by the two generative models trained on CXR-5216 and CXR-1300 datasets. The comparison includes synthetic images, the training image with the highest correlation, a scatter plot, and adjacent training slices (for brain MRI). Notably, the diffusion model's images closely resemble specific training images, indicating higher data memorization compared to StyleGAN. Each row corresponds to different model-dataset combinations.}


\begin{figure*}[h!]
  \centering
  \includegraphics[width=0.95\textwidth]{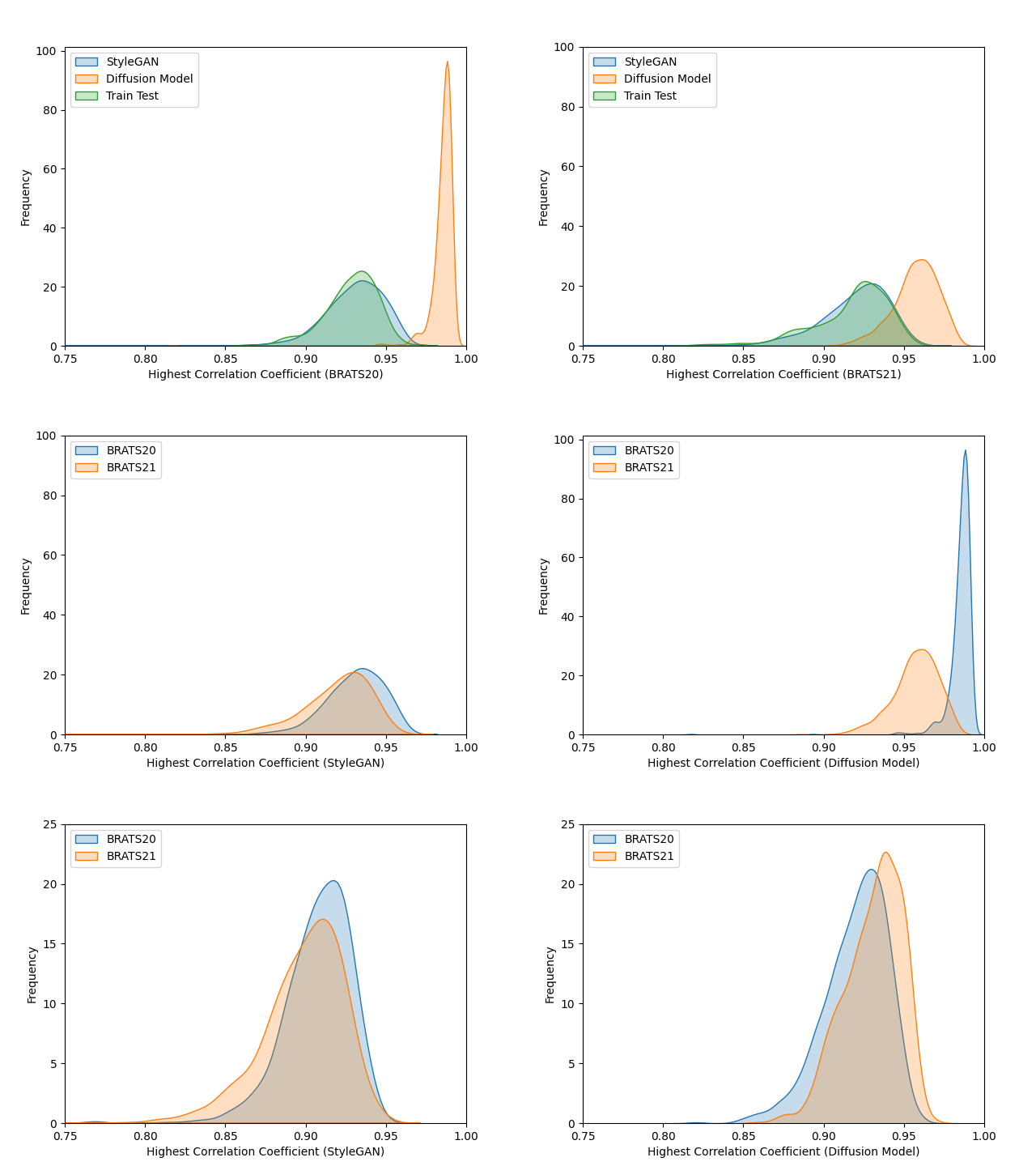}
  \caption{First row: Comparison of memorization measured as the highest correlation between \blue{1000 randomly selected synthetic images from StyleGAN and a diffusion model and all training images}. As a baseline the same comparison is done between all test images and all training images. Clearly, the diffusion model is more prone to memorization, compared to StyleGAN. Second row: A direct comparison when using BRATS20 or BRATS21 for training. Clearly, the diffusion model is more likely to memorize the training images if the training set is smaller. Third row: The same correlation analysis between synthetic images and the test images. In general the correlations are lower, but the diffusion model still produces higher correlations. 
  }
  \label{fig:Coorelation_plots}
\end{figure*}

\begin{figure*}[h!]
  \centering
  \includegraphics[width=0.95\textwidth]{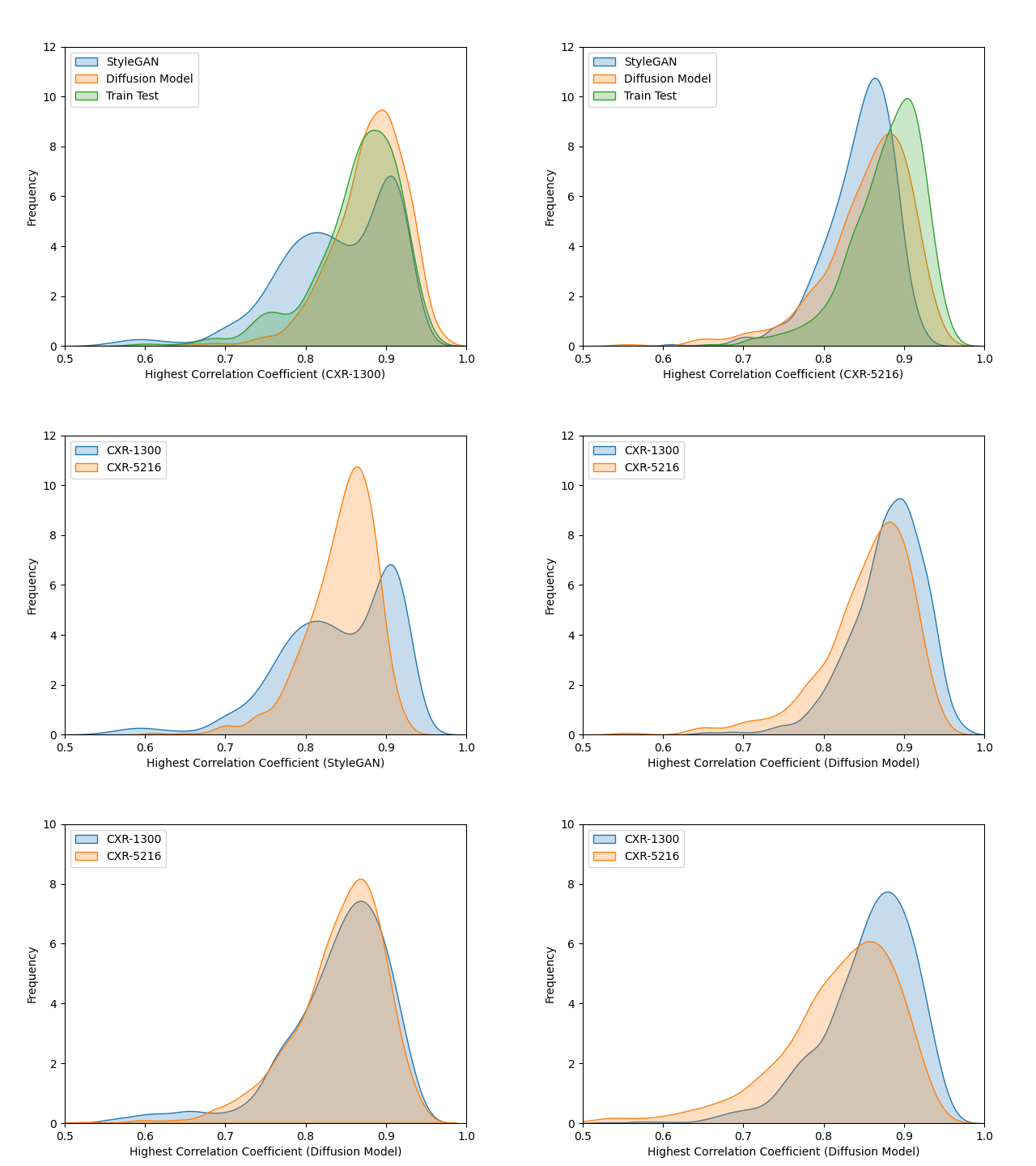} 
  \caption{First row: Comparison of memorization measured as the highest correlation between \blue{1000 randomly selected synthetic image from StyleGAN and diffusion model and all the training images.} As a baseline the same comparison is done between test images and all training images. The diffusion model is again more prone to memorization, especially for the smaller CXR1300 dataset, but the effect is not as severe as for the brain tumor images. Second row: A direct comparison when using CXR-5216 or CXR1300 for training. Both models are more likely to memorize the training images if the training set is smaller. Third row: The same correlation analysis between synthetic images and the test images. 
  }
  \label{fig:Coorelation_plots_CXR}
\end{figure*}

A total of 1000 synthetic images were randomly selected from each of our eight trainings (StyleGAN and diffusion model for BRATS20, BRATS21 and two CXR pneumonia datasets), and were used to calculate the correlation with all training images, and the highest correlations were saved. For BRATS each correlation was determined by combining all four modalities. To obtain a baseline, the correlations were also calculated between the training images and the test images from each dataset. Furthermore, the correlations were also calculated between the synthetic images and the test images. This was done to assess the level of correlation for data not used for training. From the plots in Figures~\ref{fig:Coorelation_plots} and~\ref{fig:Coorelation_plots_CXR}, it is evident that StyleGAN outperforms the diffusion model, as the latter produces images more similar to the training images. It is also clear that the degree of memorization is higher for the smaller BRATS20 dataset. Moreover from the CXR images, we can infer that the issue of memorization still persists, although not as severe. The plots demonstrate that correlation is an appropriate metric for investigating the memorization of training images. 

Table~\ref{tab:my-table} shows IS and FID for the different combinations of dataset and generative model. 
\blue{However, these metrics, including the commonly used FID, while indicative of the quality of generated images when lower, do not accurately represent the extent of data memorization relative to the training images, and we therefore also show the mean highest correlation. It is noted that generative models with better performance metrics, such as lower FID scores, often tend to memorize more data ~\cite{carlini2023extracting}, highlighting a potential discrepancy between perceived image quality and actual model generalization~\cite{carlini2023extracting}.}

\begin{table}[]
\resizebox{\columnwidth}{!}{
\begin{tabular}{|c|l|c|c|l|}
\hline
\textbf{Data}                     & \multicolumn{1}{c|}{\textbf{Model}} & \textbf{FID} & \textbf{Inception Score} & \textbf{Correlation} \\ \hline
\multirow{2}{*}{\textbf{BRATS20}} & Diffusion Model & 15.85417 & 2.35191 & 0.98510 \\ \cline{2-5} 
                                  & StyleGAN & 44.85624 & 2.34184 & 0.93037 \\ \hline
\multirow{2}{*}{\textbf{BRATS21}} & Diffusion Model & 27.54855 & 2.35156 & 0.95718 \\ \cline{2-5} 
                                  & StyleGAN & 25.34607 & 2.34360 & 0.91907 \\ \hline
\multirow{2}{*}{\textbf{CXR-5216}} & Diffusion Model & 53.81866 & 2.32937 & 0.85346 \\ \cline{2-5} 
                                  & StyleGAN & 33.14418 & 2.34287 &  0.84356 \\ \hline
\multirow{2}{*}{\textbf{CXR-1300}} & Diffusion Model  &57.45322 & 2.05646 & 0.87895 \\ \cline{2-5} 
                                  & StyleGAN        &54.68277 & 2.41764 & 0.83847 \\ \hline
\end{tabular}} 
\caption{A common comparison between GANs and diffusion models is to calculate Frechet inception distance and inception score for the synthetic images. However, these metrics do not consider memorization. Here we also show the mean highest correlation between 1000 synthetic images and all training images. Clearly the diffusion model memorizes more compared to StyleGAN.}
\label{tab:my-table}
\end{table}

\begin{table}[]
\resizebox{\columnwidth}{!}{
\begin{tabular}{|c|l|c|c|l|}
\hline
\textbf{Data}                     & \multicolumn{1}{c|}{\textbf{Model}} & \textbf{Small} & \textbf{Medium} & \textbf{Large} \\ \hline
\multirow{2}{*}{\textbf{CXR-1300}}  & Diffusion Model   & 0.86363 & 0.87895 & 0.88519 \\ \cline{2-5} 
                                    & StyleGAN          & 0.85299 & 0.83847 & 0.84336 \\ \hline
\multirow{2}{*}{\textbf{BRATS20}}   & Diffusion Model   & 0.66744* & 0.98510 & 0.97731 \\ \cline{2-5} 
                                    & StyleGAN          & 0.94756 & 0.93037 & 0.92639 \\ \hline
\end{tabular}} 
\caption{\blue{A comparison of models trained with different number of trainable parameters (medium represents the default settings). The values represent the mean highest correlation for 1000 synthetic images. The mean correlation for the small BRATS20 diffusion model is very low. This is because the model was trained for the same number of epochs as the larger one, but the images produced are extremely poor.}}
\label{tab:model_Comparison_table}
\end{table}

\blue{Table ~\ref{tab:model_Comparison_table} shows the mean highest correlation over 1000 synthetic images, for generative models with varying number of trainable parameters. The diffusion model tends to memorize more for larger models for x-ray images, while the other results are inconclusive. It should be mentioned that the synthetic images from the smaller and larger models are of lower quality compared to the medium (default) models, due to not performing an exhaustive hyperparameter search. This makes it difficult to draw any strong conclusions from this experiment. } 

\section{Discussion}

We have demonstrated that diffusion models can memorize the training images, corroborating previous results on this topic for non-medical images~\cite{somepalli2023diffusion,carlini2023extracting} as well as medical images~\cite{dar2023investigating} (here no comparison with GANs was performed). \blue{The diffusion model tends to memorize (much) more than StyleGAN, even though the number of trainable parameters is lower (depending on if one includes the trainable parameters for the GAN discriminator or not).} Memorization is problematic for medical imaging where privacy and GDPR is extremely important. Pinaya et al.~\cite{pinaya2022brain} have already shared synthetic brain volumes generated by their 3D diffusion model, and it should be investigated how similar these volumes are to the real brain volumes in UK biobank. 

\subsection{Effect of data size and data type}

We show that the degree of memorization depends on the number of training images and type of data, as the problem is less severe for BRATS21 (1195 training subjects) compared to BRATS20 (313 training subjects). We see the same effect for the pneumonia data; a smaller training set leads to more severe memorization. The synthetic images produced from BRATS exhibit a much higher correlation compared to those produced from CXR pneumonia data. \blue{One possible explanation is that the BRATS generative models have more trainable parameters, as the MR images contain 5-channels compared to a single channel for chest x-ray.} Another explanation is that BRATS contains many slices in its training set that are highly correlated with each other, as the dataset is originally 3D, while in the CXR pneumonia dataset the images are from unique subjects. Pan et al.~\cite{pan20232d}, Txurio et al.~\cite{txurio2023diffusion} and Peng et al.~\cite{peng20232d} also used 2D slices from 3D volumes, as 3D diffusion models are not yet very common. We expect that memorization is much more likely in these cases. 


\subsection{Effect of hyperparameters}

\blue{One possible explanation for the observed memorization is the number of trainable parameters each model has (in relation to the number of training images). To study how the number of trainable parameters correlate with the degree of memorization, we trained smaller versions with fewer parameters and larger versions with more parameters, for both StyleGAN and the diffusion model. The correlation results for both models are inconclusive, but indicates that more trainable parameters can lead to more memorization for diffusion models. Training GANs and diffusion models is very time consuming due to the complexity of their architecture and the iterative nature of their training process. Given these constraints, conducting numerous experiments to fine-tune all hyperparameters becomes impractical, and we therefore only changed a few hyperparameters to make the models smaller or larger. This resulted in artefacts in the synthetic images, making it difficult to draw strong conclusions regarding how the number of trainable parameters affect memorization.}

\subsection{Importance of detecting memorization}

Medical images often contain identifiable information that can be linked to specific individuals, making handling and storage of this data a matter of significant concern. Mismanagement or unauthorized access to such data could lead to breaches of patient privacy, potentially resulting in legal consequences and reputational damage for medical institutions. Therefore, it is essential to implement strict security measures and adhere to privacy regulations when working with medical imaging data. In the majority of cases, medical data is deemed sensitive and is often restricted from being shared, even for research purposes. This is due to various concerns, including ethical considerations, privacy protection, and the sensitive nature of the information itself. Using synthetic images can resolve many issues and may be considered safe for sharing, as synthetic images do not belong to a specific person. However, a concern remains: what if the models generating these images are not creating unique content, but instead reproduce the training data?


For big generative models, one can wonder if they create genuinely new results or if they just copy and mix elements from their training data. By examining how these models remember, it can be understood how much they copy from the data they were trained on. Saharia et al.~\cite{saharia2022photorealistic} found that overfitting is not a problem in the model they trained model, and that more training could improve the performance of the diffusion model. This leads us to question if diffusion models perform better than previous methods because they remember more information.

\subsection{Preventing memorization}

There are several potential ways to prevent memorization when training generative models. \blue{For example, one may introduce a memorization penalty in the loss function~\cite{borji2022pros}}. Another approach is to only use augmented images (i.e. after applying rotation, flipping, elastic deformations, etc), and to never use the original images. However, this will make it harder and much more time consuming to evaluate how similar the synthetic images are to the original training images, if it is necessary to first apply image registration to be able to compare each image pair. Similar to Dar et al.~\cite{dar2023investigating}, one can train an autoencoder to compress each image or volume, and instead measure the similarity of the feature vectors. If the autoencoder is trained with augmentation, it should be able to detect a high similarity even if two images have different orientation. \blue{As previously mentioned, the number of elements in the feature vector may however be too low to detect memorization of fine details. As proposed in our related work~\cite{akbar2023brain}, to fine-tune generative models pre-trained on large open datasets can potentially lead to less memorization, compared to training these models from scratch.} Another approach for preventing memorization is to use techniques similar to differential privacy (DP)~\cite{abadi2016deep}, but recent work demonstrated that diffusion models trained with DP did not converge~\cite{carlini2023extracting}. To memorize the training images is an overfitting problem, and common regularization techniques such as dropout and weight decay can potentially help. Recent work has demonstrated that model gradient similarity~\cite{szolnoky2022interpretability} can be very helpful to automatically penalize memorization.

\section{Conclusion}

This study highlights an important issue with using diffusion models \blue{(and to some extent GANs)} for medical images: the problem of these models remembering and possibly copying the data they were trained on. This is especially true for small datasets or datasets with very similar images. This not only causes problems with how well the models work, but also raises privacy concerns because there's a chance these models might recreate images that can be linked to real sensitive data. \blue{We have here studied memorization for brain and chest images, but the problem is equally important for all types of medical images.}

We found that memorization is more noticeable in certain datasets, like BRATS, than in others like the CXR pneumonia dataset which has more varied images. This tells us that how a model behaves depends a lot on the kind of data it learns from and how much data there is.
Considering these points, it is really important for people who make and use these models, especially in medical imaging, to find the right balance. They need to use the models' power for good while making sure they handle private medical information correctly. Looking ahead, we should keep looking for ways to prevent these models from memorizing too much and think carefully about the ethical side of using them in healthcare.

\section{Acknowledgements}

Training StyleGAN, and sampling from the diffusion models, was performed using the supercomputing resource Berzelius (752 Nvidia A100 GPUs) provided by the National Supercomputer Centre at Linköping University, Sweden. It was donated by the Knut and Alice Wallenberg Foundation. 

This research was supported by the ITEA/VINNOVA project ASSIST (Automation, Surgery Support and Intuitive 3D visualization to optimize workflow in IGT SysTems, 2021-01954), LiU Cancer and the Åke Wiberg foundation. Anders Eklund has previously received graphics hardware from Nvidia.

\printbibliography

\end{document}